\documentclass{aa}  
\usepackage{graphicx}
\usepackage{txfonts}
\begin{document}

   \title{Propagating Kink Waves in an Open Coronal Magnetic Flux Tube with Gravitational Stratification: Magnetohydrodynamic Simulation and Forward Modelling}


   \author{Yuhang Gao
          \inst{1, 2},  
          Tom Van Doorsselaere
          \inst{2}, 
          Hui Tian
          \inst{1},
          Mingzhe Guo\inst{2,3},
          \and
          Konstantinos Karampelas\inst{2}
          }
         \institute{School of Earth and Space Sciences, Peking University, Beijing, 100871, People's Republic of China\\
         \email{huitian@pku.edu.cn, yh\_gao@pku.edu.cn}
         \and
         Centre for mathematical Plasma-Astrophysics, Department of Mathematics, KU Leuven, Celestijnenlaan 200B bus 2400, B-3001 Leuven, Belgium
         \and
         Shandong Provincial Key Laboratory of Optical Astronomy and Solar-Terrestrial Environment, Institute of Space Sciences, Shandong University, Weihai 264209, China
             }

   \date{Received ; accepted }

 
  \abstract
   {In the coronal open-field regions, such as coronal holes, there are many transverse waves propagating along magnetic flux tubes, generally interpreted as kink waves. Previous studies have highlighted their potential in coronal heating, solar wind acceleration, and seismological diagnostics of various physical parameters.}
   {This study aims to investigate propagating kink waves, considering both vertical and horizontal density inhomogeneity, using three-dimensional magnetohydrodynamic (MHD) simulations.}
   {We establish a 3D MHD model of a gravitationally stratified open flux tube, incorporating a velocity driver at the lower boundary to excite propagating kink waves. Forward modelling is conducted to synthesise observational signatures of the Fe \textsc{ix} 17.1 nm line.}
   {It is found that resonant absorption and density stratification both affect the wave amplitude. When diagnosing the relative density profile with velocity amplitude, resonant damping needs to be properly considered to avoid possible underestimation. In addition, unlike standing modes, propagating waves are believed to be Kelvin-Helmholtz stable. In the presence of vertical stratification, however, phase mixing of transverse motions around the tube boundary can still induce small scales, partially dissipating wave energy and leading to a temperature increase, especially at higher altitudes. Moreover, forward modeling is conducted to synthesise observational signatures, revealing the promising potential of future coronal imaging spectrometers such as MUSE in resolving these wave-induced signatures. Also, the synthesised intensity signals exhibit apparent periodic variations, offering a potential method to indirectly observe propagating kink waves with current EUV imagers.}
   {}

   \keywords{solar atmosphere --
                solar corona --
                solar oscillations --
                magnetohydrodynamical simulations 
               }

   \titlerunning{Propagating Kink waves in a stratified open flux tube}
   \authorrunning{Y. Gao et al.}

   \maketitle
%
\section{Introduction} \label{sec:intro}

In the past two decades, propagating transverse motions have been frequently detected in the solar corona \citep[see the reviews by][]{Banerjee2021,morton2023}. These waves are commonly interpreted as kink waves \citep{tvd2008,goossens2009}. Observations reveal their ubiquitous presence in the solar corona \citep[e.g.,][]{tomczyk2007,tomczyk2009}, carrying substantial energy flux that could potentially contribute to coronal heating and solar wind acceleration \citep[e.g.,][]{banerjee2009, Hahn2012,McIntosh2011}. Another important application of these waves is in coronal seismology. For instance, with the propagating kink waves observed by the Coronal Multichannel Polarimeter \citep[CoMP;][]{tomczyk2008}, global maps of the coronal magnetic field can be obtained \citep[][]{yang2020,yang2020Sci}, marking a significant advancement toward routine measurement of the coronal magnetic field. Therefore, the propagating kink waves are of great significance in solar physics research.

In general, kink waves primarily manifest as propagating modes in open-field regions or some large closed magnetic structures (i.e., coronal loops), while exhibiting as standing modes in relatively smaller coronal loops in both quiet-Sun and active regions \citep[][]{naka1999,aschwanden1999,tian2012,nistico2013,anfin2015,goddard2016,gao2022,li2023,petrova2023}{}{}. It has not been fully understood how and why these transverse waves appear differently in coronal loops of varying sizes (see e.g., \citealt{Threlfall2013,long2017,morton2021,tiwari2021,skirvin2023,gao2023,li2023traveling}
). In this paper, we focus on kink waves in open-field coronal regions, for example, coronal holes, where the propagating modes naturally predominate. The typical structures or waveguides here are plumes, characterised by a thin, long ray-like appearance \citep[see the review by][]{Poletto2015}. 

In observations, kink waves propagating along plumes or coronal open flux tubes can be studied using spectroscopic observations or extreme ultraviolet (EUV) image sequences. On one hand, spectroscopic observations are mainly based on the CoMP Fe \textsc{xiii} 1074.7 nm emission line, which can show the existence of persistent Doppler velocity fluctuations distributing nearly everywhere in the off-limb corona \citep{tomczyk2007,tomczyk2009,yang2020,yang2020Sci}. Based on the wave-tracking method, the propagating speed or phase speed was estimated to be 300-700 km/s. In the open-field coronal hole regions, the Fe \textsc{xiii} 1074.7 nm line normally has a low signal-to-noise ratio. Nevertheless, some authors still managed to detect signals of propagating kink motions with a velocity amplitude of several kilometers per second \citep{liu2015,morton2015}. Similar signatures were also captured in \cite{mancuso2015} with H \textsc{i} Ly $\alpha$ coronal emission line profiles. \cite{morton2015} also revealed the existence of downward propagating waves, with a decreased power compared with that of upward propagating ones. On the other hand, EUV imaging observations with a high spatial and temporal resolution provide us with another approach to investigate transverse waves in open-field regions through the plane-of-sky (POS) displacements of magnetised plasma structures like plumes \citep{McIntosh2011,thurgood2014,morton2015,weberg2018,weberg2020}. Most previous EUV observations used the data from the Atmospheric Imaging Assembly \citep[AIA;][]{lemen2012} on board the Solar Dynamics Observatory \citep[SDO;][]{pesnell2012}. Some comparisons between imaging and spectroscopic observation results can be found in \cite{morton2015,morton2019}. Both observations show a peak at $\sim3.5$ mHz in the velocity power spectrum \citep[see Figure 1(d) in][]{morton2019}, suggesting a connection with photospheric p-mode leakage \citep[see also][]{tomczyk2009,weberg2020,gao2022,gao2023}.
Spectroscopic analysis has also shown the variation of non-thermal line broadening, which could be interpreted as the signature of Alfv\'{e}n waves \citep[e.g.,][]{Banerjee1998,banerjee2009, Doschek2007,jess2009, Hahn2012}. However, such an interpretation remains under debate, because the line broadening can be attributed to other effects such as flow inhomogeneities along the line of sight or sausage modes \citep[see e.g.,][]{tian2011,tian2014,Mathioudakis2013,zhu2023}. 
The direct observation of torsional Alfv\'{e}n waves in the coronal open-field regions is still lacking \citep[c.f.,][]{kohutova2020,Petrova2024}. Therefore, here we restricted the observed transverse waves to kink waves only \citep[see][]{tvd2008,goossens2009,goossens2012}.

The propagation of kink waves in the magnetised, inhomogeneous coronal medium is influenced by two primary effects. Firstly, the amplitude increases with height due to decreasing density. Assuming that the wave damping or dissipation is relatively weak,
the energy flux $F=\langle\rho\rangle v_\mathrm{amp}^2c_\mathrm{k}$ would be nearly constant, where $\langle\rho\rangle$ represents the average density, $v_\mathrm{amp}$ denotes the velocity amplitude, and $c_\mathrm{k}$ stands for local kink speed (phase speed), defined as $c_\mathrm{k}(z)=B(z)/\sqrt{\mu_0 \langle \rho\rangle}$ with $\mu_0$ representing the magnetic permeability \citep[e.g.,][]{tomczyk2009,morton2015,yang2020Sci,bate2022}. Thus, if the magnetic field $B(z)$ only changes slightly, the decrease in density with height can result in an amplification of wave amplitude  \citep{Banerjee1998,banerjee2009,soler2011,morton2012,morton2014,vanBallegooijen2014,weberg2020}. The second effect involves the resonant damping of the kink waves \citep[e.g.,][]{terradas2010, verth2010,pascoe2010,pascoe2011,pascoe2012}. The damping arises from the inhomogeneity across the loop, transferring energy from the bulk transverse oscillations to azimuthal motions in the boundary layer. Subsequently, the energy can be dissipated through phase mixing \citep[][]{Heyvaerts1983,soler2015,Pagano2017,howson2020} and the development of Kelvin-Helmholtz instability \citep[KHI; see][]{pagano2019}. Other effects such as uniturbulence may also contribute to wave dissipation \citep[][]{magyar2017,magyar2019,tvd2020}.

While the joint influence of these two effects on propagating kink waves has been analytically investigated in \cite{soler2011}, a detailed study with 3D magnetohydrodynamic (MHD) simulation considering both vertical and horizontal inhomogeneities is still very rare. A few previous numerical investigations \citep{pascoe2010,pascoe2011,pascoe2012, DeMoortel2012,pascoe2015,deMoortel2016,magyar2017, Pagano2017,pagano2019,pagano2020,howson2020, Fyfe2021, Meringolo2023} have been focusing on the damping mechanism and heating contribution of these waves without considering vertical density stratification, which can be important as it can amplify the amplitude and thus affect the wave damping/dissipation. The varying density also influences the estimation of wave energy flux and total radiative loss. On the other hand, studies incorporating density variation along the magnetic field usually neglect the density inhomogeneity across the field \citep[e.g.,][]{vanBallegooijen2014, Matsumoto2018,pant2020,pascoe2022}, and consequently, the resonant absorption and phase mixing are not existing or effective. Recently, \cite{pant2019} and \cite{sen2021} established models with multiple density-enhanced flux tubes and gravitational stratification. However, after kink waves are excited, these oscillating tubes quickly interact with each other and generate (uni)turbulence, making it impossible to investigate the wave behaviour of one single flux tube in detail. Additionally, \cite{pelouze2023} built up a single flux tube model with stratification, but only focusing on the cut-off effect at the lower atmosphere rather than the coronal part.
 

In this study, we report results from a detailed investigation of the propagating kink waves in open coronal magnetic flux tubes (or plumes) perpendicular to the solar surface with 3D MHD simulations. Both horizontal inhomogeneity and vertical stratification are considered to better reflect the realistic case. We also perform forward modelling to synthesise observational signatures of these waves. The paper is organised as follows: In Section \ref{sec:method}, we describe our simulation setup and methodology. Section \ref{sec:res} and \ref{sec:fomo} present the simulation and forward-modelling results, respectively. We further discuss our results in Section \ref{sec:discussion} and summarise our findings in Section \ref{sec:conclusion}.

\begin{figure*}
    \centering
    \includegraphics[width=1.0\textwidth]{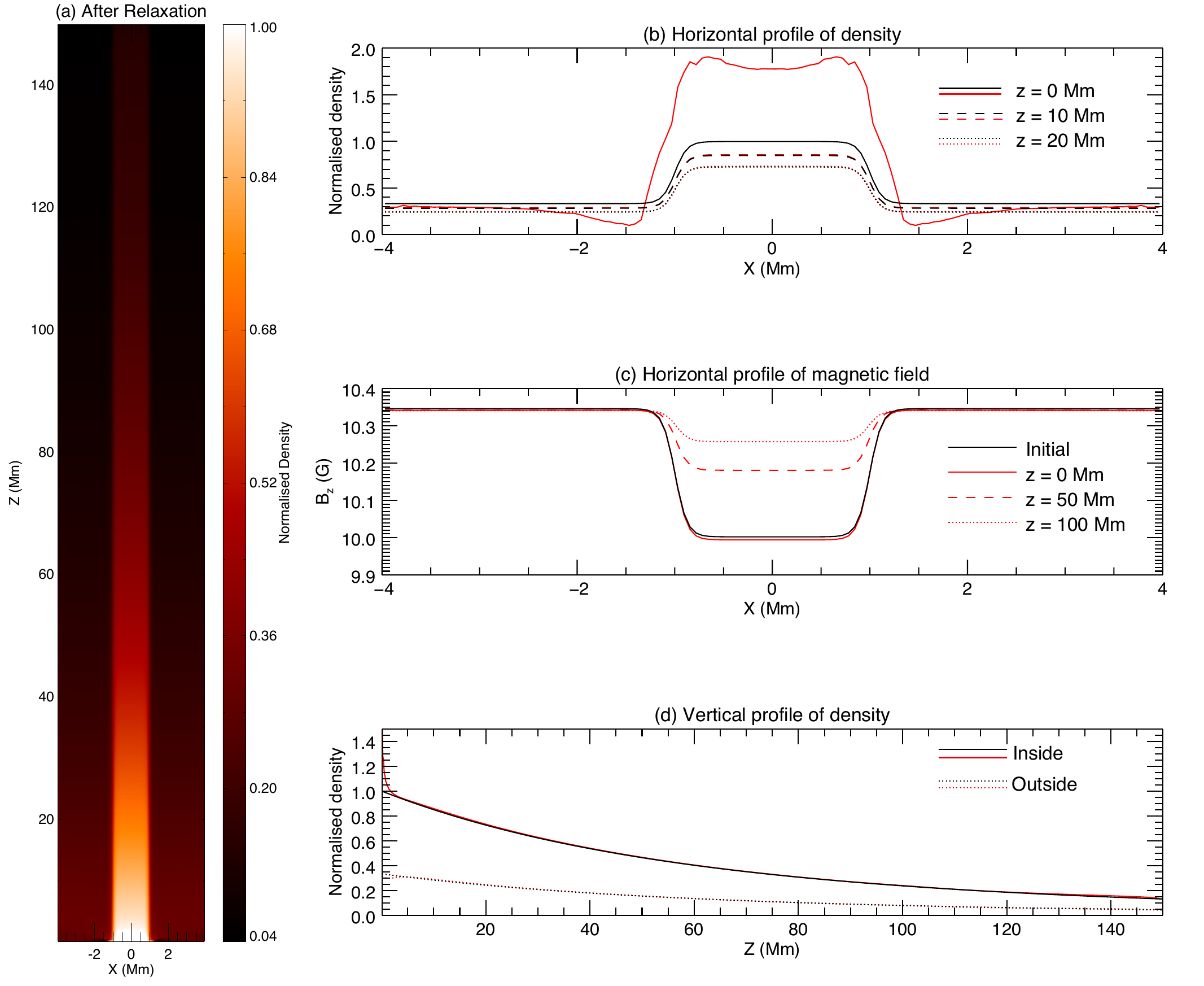}
    \caption{Illustration of the model setup. (a) Density distribution at the $y=0$ plane after relaxation. (b)-(c) Horizontal profiles of density and magnetic field along the $x$ axis at three different heights (indicated by different line styles). (d) Vertical profile of density demonstrating gravitational stratification, with solid and dashed lines indicating density within and outside the flux tube, respectively. Black and red lines in panels (b)-(d) represent states before and after relaxation, respectively. We note that the magnetic field at the initial state (before relaxation) is uniform along the $z$ axis, hence there is only one black line in panel (c).}
    \label{fig:1}
\end{figure*}

\section{Method}\label{sec:method}

\begin{figure}
    \centering
    \includegraphics[width=0.45\textwidth]{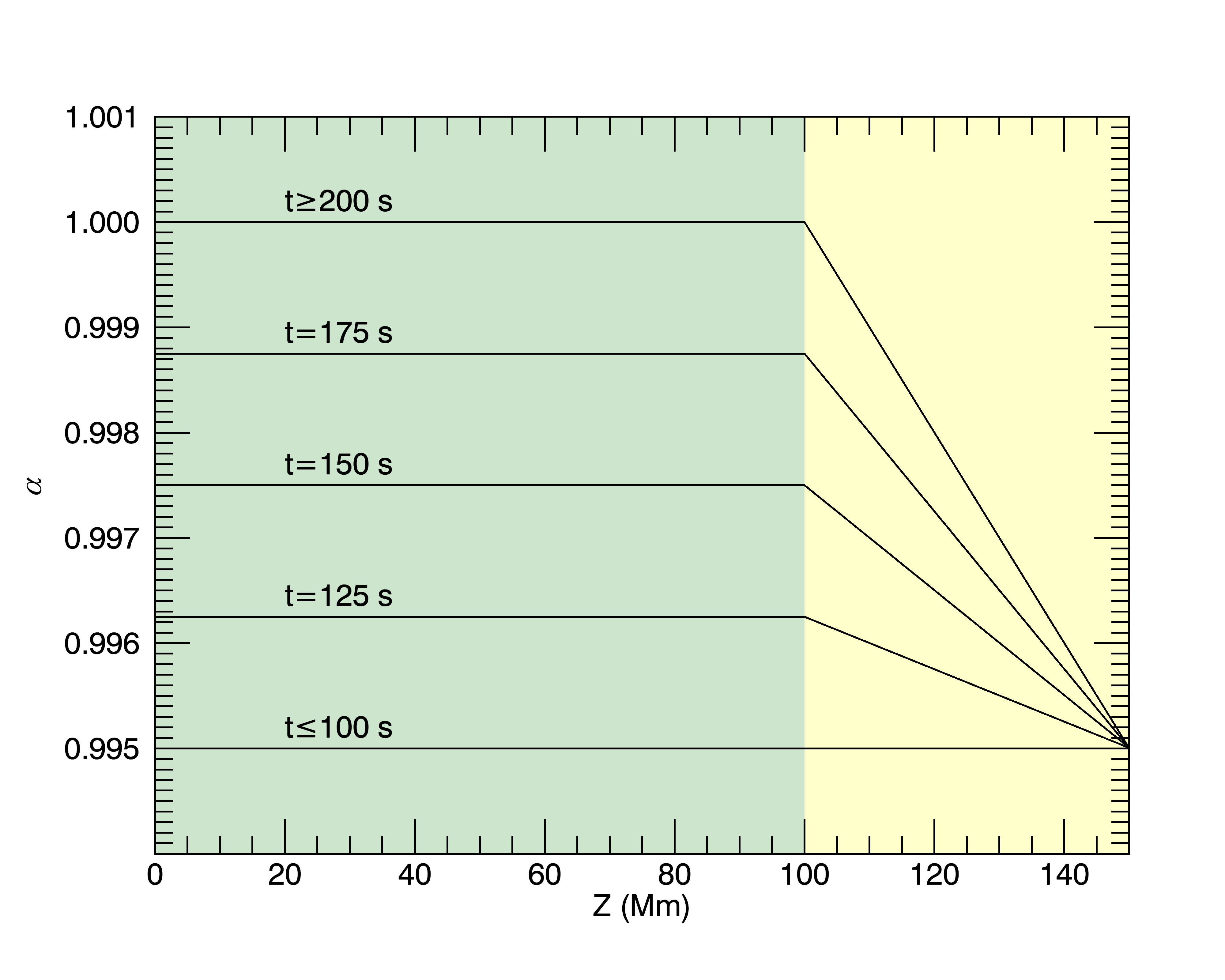}
    \caption{Dependence of the velocity absorption factor $\alpha$ on height ($z$, horizontal axis) and time (indicated by different lines). Green and yellow shading distinguishes the regions below and above 100 Mm. The latter can be regarded as an extended upper boundary, and we only analyse and discuss the results below 100 Mm (green shading).}
    \label{fig:2}
\end{figure}

In this study, we aim to model a plume in the coronal open field region as a magnetic flux tube in 3D MHD simulations. 
The length $L$ of the tube is chosen to be 150 Mm along the $z$ axis in Cartesian coordinates. 
We include a gravity along the tube specified as
\begin{equation}
    \vec{g}=-\frac{g_0R_\odot^2}{(R_\odot+z)^2}\Vec{e}_z\,,
\end{equation}
to reflect the stratification, where $R_\odot$ is the solar radius, $g_0=274\,\text{m s}^{-2}$ is the gravity at the solar surface (or, approximately, at the base of the corona). The lower solar atmosphere is not considered in our model. Thus, we adopt a uniform temperature as $T=1$ MK.

At the lower boundary, the horizontal density profile is set as
\begin{equation}
    \rho(x,y,z=0)=\rho_\text{e}+\frac{(\rho_\text{i}-\rho_\text{e})}{2}\left\{1-\tanh\left[\left(\frac{\sqrt{x^2+y^2}}{R}-1\right)b\right]\right\}\,.
\end{equation}
The external and internal densities are set to be $\rho_\mathrm{e}=2.5\times10^{-15}\,\text{g cm}^{-3}$, and $\rho_\mathrm{i}=7.5\times10^{-15}\,\text{g cm}^{-3}$, respectively, resulting in a density contrast ($\zeta=\rho_\mathrm{i}/\rho_\mathrm{e})$ of 3 \citep[e.g.,][]{verwichte2013,morton2021}. The tube radius $R$ and the dimensionless constant $b$ are set to be 1 Mm and 8, resulting in a boundary layer with a width of $l\sim0.6R=0.6$ Mm. The density at larger heights is integrated layer by layer to satisfy the field-aligned hydrostatic equilibrium with a finite difference scheme as described in \cite{pelouze2023,guo2023conduction,gao2023} and \cite{kara2024}. The horizontal and vertical profiles of density at the initial state are shown with black lines in Figure \ref{fig:1}(b) and (d). 

The magnetic field is initially set to maintain total pressure balance at the lower boundary, with an internal strength of 10 G, as shown in Figure \ref{fig:1}(c). Constrained only in the $z$ direction, the magnetic field exhibits zero gradients along the $z$ axis to ensure $\nabla\cdot\vec{B}=0$. Such an initial setup is not in magnetohydrostatic (MHS) equilibrium, thus a relaxation for 2400\,s is conducted to reach total pressure balance at every height \citep[see][]{pelouze2023,guo2023conduction,gao2023,kara2024}. During the relaxation stage, we introduce a velocity absorption factor $\alpha$ to suppress initial velocity flows in the simulation domain and reflections from the upper boundary, defined as
\begin{equation}
    \alpha(z,t)=\left\{
    \begin{aligned}
        & 0.995, & t\le t_{0},\\
        & \left(0.995+0.005\times\frac{t-t_0}{t_0}\right)\left[1-(1-\alpha^*(z))\frac{t-t_0}{t_0}\right], & t_{0}<t\le 2t_{0},\\
        & \alpha^*(z),& t> 2t_{0}.
    \end{aligned}
    \right.
\end{equation}
Here $\alpha^*(z)$ describes the absorption factor at the final stage:
\begin{equation}
    \alpha^*(z)=\left\{
    \begin{aligned}
    & 1, & z\le L_0,\\
    & 1-0.005\times\frac{z-L_0}{L-L_0}, & z> L_0,\\
    \end{aligned}
    \right.
\end{equation}
The velocity is multiplied by $\alpha(z,t)\le1$, and we set $t_0$ and $L_0$ to be 100 s and 100 Mm. The introduction of the factor is inspired by some previous studies \citep{pelouze2023,guo2023conduction,gao2023}. The main difference here is that we set the upper 50 Mm ($z>L_0=100$ Mm) as a permanent absorption layer to continuously suppress the reflection from the upper boundary (even after the relaxation stage). 
Figure \ref{fig:2} illustrates the distribution and evolution of $\alpha$.

After the relaxation, the horizontal velocities $v_x$ and $v_y$ remain negligible with a maximum value of 0.03 km\,s$^{-1}$, while the vertical velocity $v_z$ can be gradually suppressed to less than 1.5 km\,s$^{-1}$, much smaller than the velocity of the driven kink waves. Therefore, we regard the relaxed state (achieved after 2400 s) as a quasi-MHS equilibrium. As shown in Figure \ref{fig:1} (b)-(d), the distributions of density and magnetic field only exhibit minor changes after relaxation (indicated by the red lines). The primary density change occurs near the base of the flux tube. A possible reason is that the vertical hydrostatic equilibrium in our initial setup is not completely accurate, resulting in a net downward mass flow. 
Additionally, the magnetic field inside the flux tube experiences a slight increase with height, which is in line with the expectation that at higher altitudes, the difference in internal and external thermal pressure diminishes, resulting in a decrease in the magnetic pressure difference. Furthermore, this change in the magnetic field can lead to an increase in magnetic pressure with height, creating an upward vertical gradient in the total pressure, which could also cause density to accumulate at the tube base. Finally, Figure \ref{fig:1}(a) provides a direct visualisation of our model (post-relaxation) by illustrating the density profile at the $y=0$ plane.

To excite propagating kink waves, we add a continuous velocity driver at the lower boundary ($z=0$). The driver has a dipole-like velocity field, similar to \cite{pascoe2010,kara2017} and \cite{guo2019}. The velocity inside the tube has only the $x$-component, given by
\begin{equation}\label{eq:driver}
    \vec{v}(x,y)=v_0\cos\left(\frac{2\pi t}{P}\right)\vec{e}_x\,;
\end{equation}
and the velocity outside the tube is given by
\begin{equation}\label{eq:driver}
    \vec{v}(x,y)=v_0R^2\cos\left(\frac{2\pi t}{P}\right)\left[\frac{x^2-y^2}{(x^2+y^2)^2}\vec{e}_x+\frac{2xy}{(x^2+y^2)^2}\vec{e}_y\right]\,,
\end{equation}
where we adopt a velocity amplitude ($v_0$) of 8 km/s, and a period ($P$) of 300 s \citep[see e.g.,][]{tomczyk2009,weberg2018}. 
For simplicity, here we only use a monoperiodic driver. We note that the propagating kink waves can be better simulated with a broad-band or multi-frequency driver \citep[e.g.,][]{pascoe2015,magyar2017,magyar2018,pagano2019,pant2019,sen2021}. However, such a simplified driver can already satisfy the main purpose of this study, and the case with a broad-band driver will be explored in a future study.

The PLUTO code \citep{mignone2007} is utilised to solve the 3D time-dependent non-linear ideal MHD equations. We employ a second-order parabolic spatial scheme for time stepping, and calculate numerical fluxes with a Roe Riemann solver. The radiative cooling, thermal conduction, explicit resistivity, and viscosity are not included during the simulation. We adopt the hyperbolic divergence cleaning method to maintain the divergence-free nature of the magnetic field. 
We choose not to use dimensionless parameters because using dimensional parameters allows for more intuitive comparisons with real observations.
The computational domain spans [-4, 4] Mm $\times$ [-4, 4] Mm $\times$ [0, 150] Mm, with a uniform grid of
$128\times128\times1024$ cells, yielding resolutions of 62.5 km in the $x$ and $y$ directions, and 146.5 km in the $z$ direction. 
Although the horizontal resolution could be relatively lower compared to previous studies, the small-scale structures associated with resonant absorption and phase mixing can still form naturally in the tube boundary (see below).
Such a low resolution also induces a large numerical resistivity $\eta_\mathrm{n}$, which is roughly estimated to be of the order of $10^{-7}$ s (in CGS), further giving a magnetic Reynolds number of $\sim2.8\times10^3$. Considering that the numerical resistivity is much larger than the expected value of the solar corona and the magnetic Reynolds number is much lower \citep[see][]{kara2017}, we perform a convergence test by re-running the simulation with double
horizontal resolution. It is found that the simulation results such as temperature increase (see Section \ref{sec:res}) only show mild changes without affecting our main conclusions.

For the boundary condition, we employ outflow conditions for all the lateral boundaries and the upper boundary ($z=150$ Mm). During the relaxation stage, the lower boundary ($z=0$) remains fixed with all velocities set to zero. Subsequently, we modify the velocity conditions according to Equation (\ref{eq:driver}) to drive upward-propagating kink waves. We also note that an outflow upper boundary cannot fully eliminate the reflective flows/waves, so we keep the upper 50 Mm ($z>100$ Mm) as a velocity absorption layer or an extended upper boundary throughout the whole simulation (as shown in Figure \ref{fig:2}), to further diminish the reflection \citep[see also][]{pelouze2023}. Given that, when analysing the results, we only pay attention to the part of $z\le100$ Mm.

\section{Results}\label{sec:res}

\subsection{Height dependence of the wave amplitude}\label{subsec:amplitude}

\begin{figure*}
    \centering
    \includegraphics[width=1.0\textwidth]{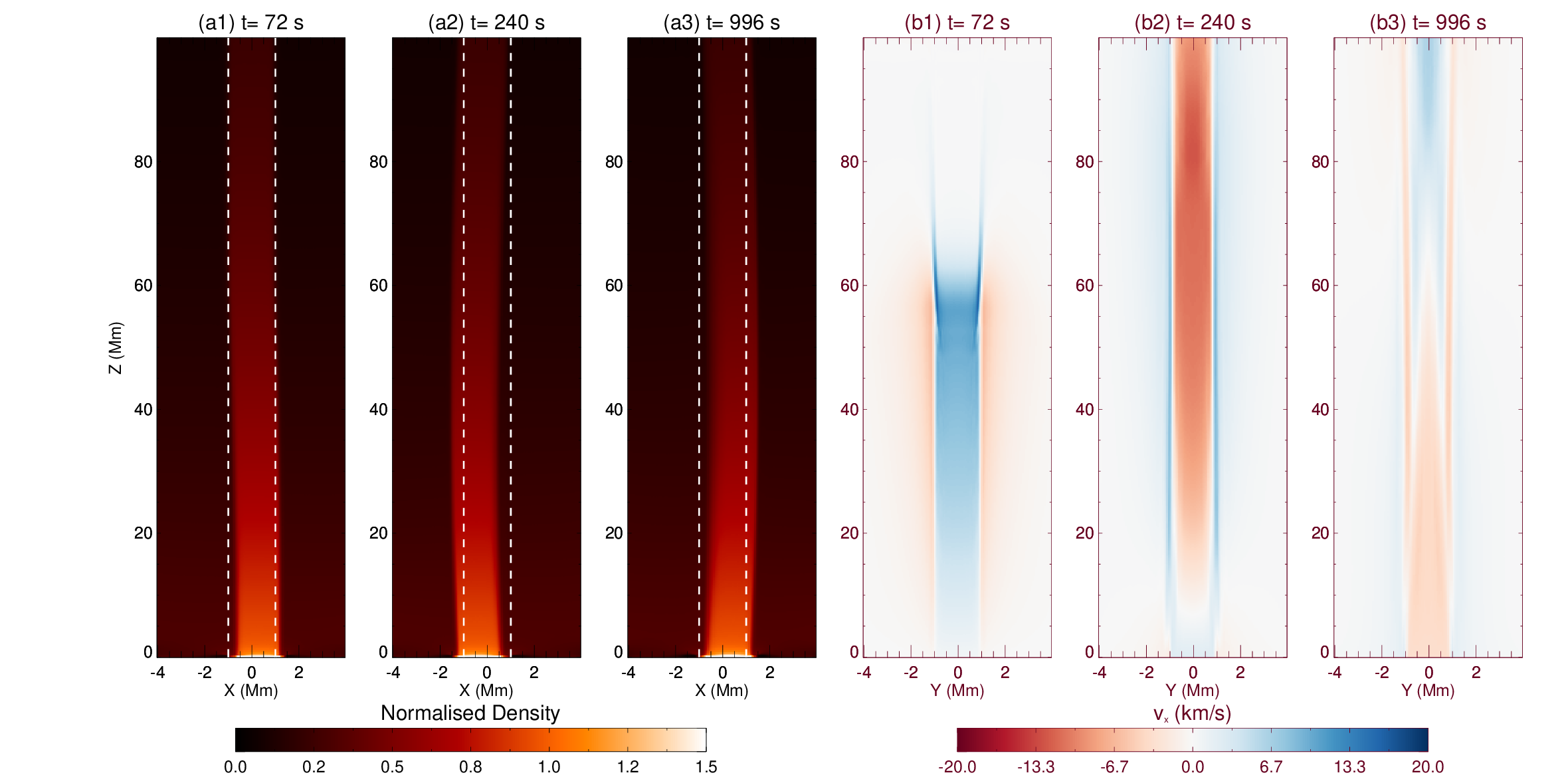}
    \caption{Snapshots at three different time steps ($t=72$ s, 240 s, and 996 s) after propagating kink waves are excited. (a1)-(a3) Normalised density (normalised with the code unit $10^9$ cm$^-3$) at $y=0$ plane. The dashed white lines indicate the unperturbed tube boundary (i.e., $x=\pm 1$ Mm). (b1)-(b3) $v_x$ at $x=0$ plane. Relevant animations are available in the online Journal.}
    \label{fig:3}
\end{figure*}

\begin{figure*}
    \centering
    \includegraphics[width=1.0\textwidth]{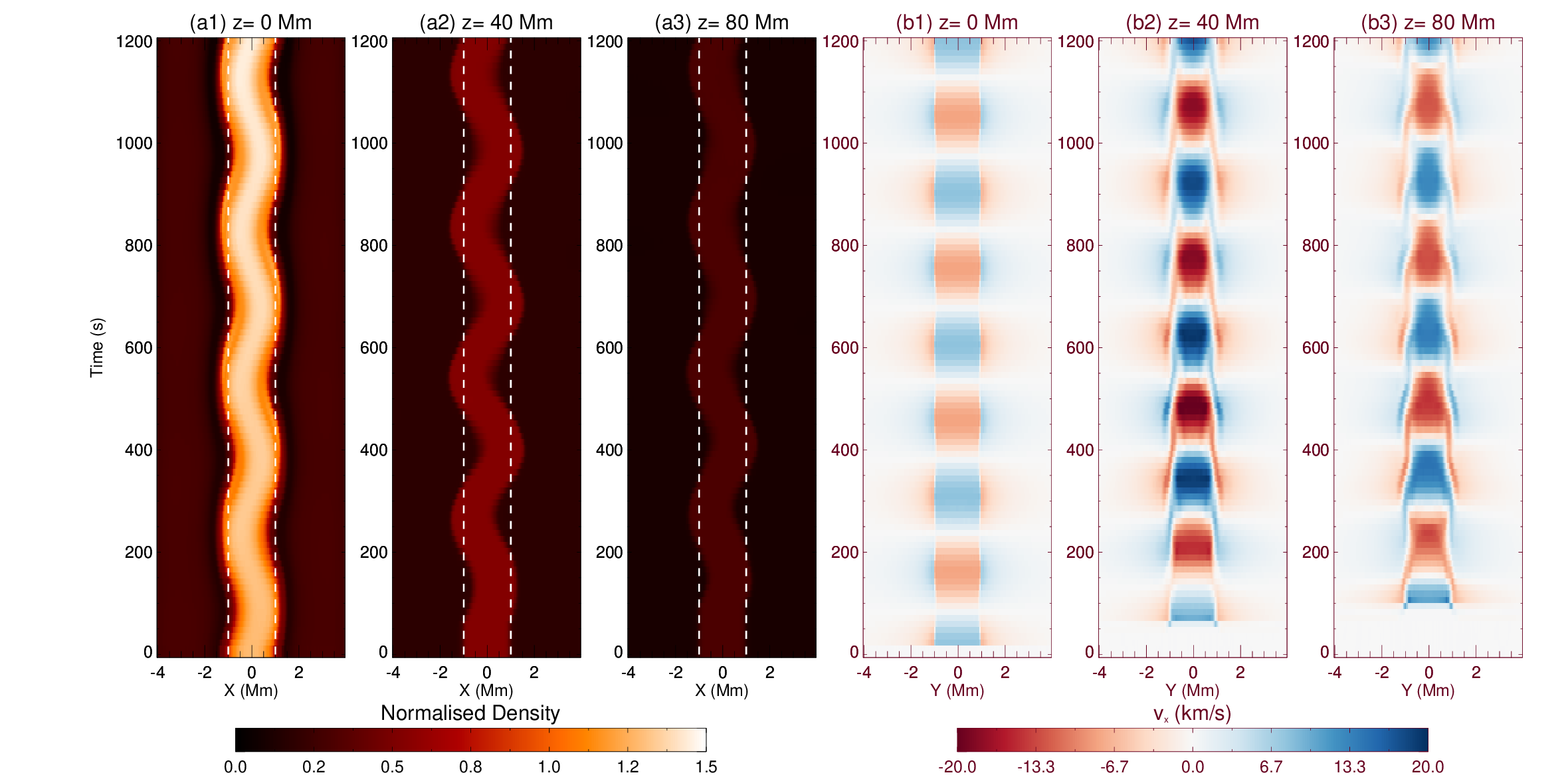}
    \caption{Time-distance maps at three different heights ($z=0$ Mm, 40 Mm, and 80 Mm). (a1)-(a3) Normalised density at $y=0$. The dashed white lines indicate the unperturbed tube boundary (i.e., $x=\pm 1$ Mm). (b1)-(b3) $v_x$ at $x=0$.}
    \label{fig:4}
\end{figure*}

After the velocity driver is added, propagating kink waves are excited. The velocity perturbations will propagate upwards at the local kink speed. In Figure \ref{fig:3}(a), we present the state of the flux tube by plotting density at the $y=0$ plane at three different times: $t=72$ s, 240 s, and 996 s. The distribution of corresponding transverse velocities $v_x$ are illustrated in Figure \ref{fig:3}(b). We plot them in the $x=0$ plane, thus the velocity signals can be seen as the line-of-sight (LOS) velocity. At $t=996$ s, we can already notice that some small scales have formed at the tube boundary due to phase mixing \citep[see also][]{pascoe2010,guo2020}. 

The evolution of the flux tube is further illustrated with time-distance (TD) maps in Figure \ref{fig:4}. The TD maps of density at three different heights all show sinusoidal transverse displacements, which is a typical characteristic of kink waves.
The effect of density stratification and resonant absorption can be seen in panels (b1)-(b3), which present TD maps of LOS velocity (i.e., $v_x$ at $x=0$ plane). The velocity amplitude at the tube centre (inside the inhomogeneous boundary layer) represents the energy of driven bulk kink waves. It first increases from $z=0$ Mm to 40 Mm and then decreases from 40 Mm to 80 Mm. The former is due to the density decreasing with height, as a result of gravity. However, the resonant absorption is already in effect, which can be inferred from the formation of fine structures at the boundary layer. As indicated in e.g., \cite{terradas2010,pascoe2010} and \cite{soler2011}, the energy of kink waves transfers to azimuthal Alfv\'{e}n modes in the inhomogeneous layer, leading to a coupling of kink mode and Alfv\'{e}n mode. The effect of resonant absorption becomes more dominant as the height increases, because at every height, the transverse motions (in a coupling state of kink and Alfv\'{e}n modes) act as a new wave source. Then the coupling state is inherited by waves at higher heights, and resonant absorption can develop based on this state \citep[e.g.,][]{pascoe2011,pascoe2013,hood2013,deMoortel2022}. In fact, if we constrain the driver to only one period, the kink wave in the tube centre will decay very quickly and soon there will be only torsional Alfv\'{e}n waves at the tube boundary, as shown in \cite{pascoe2010,pascoe2012} and \cite{Pagano2017}. In contrast, in our case with a continuous driver, the kink wave will not be fully damped, and its amplitude can even be amplified due to density stratification. Nevertheless, resonant absorption dominates at higher heights, leading to a reduced amplitude at $z=80$ Mm. We can further obtain the velocity amplitude at each height ($z_0$) by fitting the time series of $v_x(0,0,z_0)$ (see below). The amplitude reaches its peak at $z\sim45$ Mm. Below that height, the stratification-induced amplitude increase is more dominant; while over that height, the resonant damping becomes more effective. It is consistent with the statement that the amplitude is simply a product of these two effects given in \cite{soler2011}.

\subsection{Development of small-scale structures and relevant heating}\label{subsec:heating}

\begin{figure}
    \centering
    \includegraphics[width=0.45\textwidth]{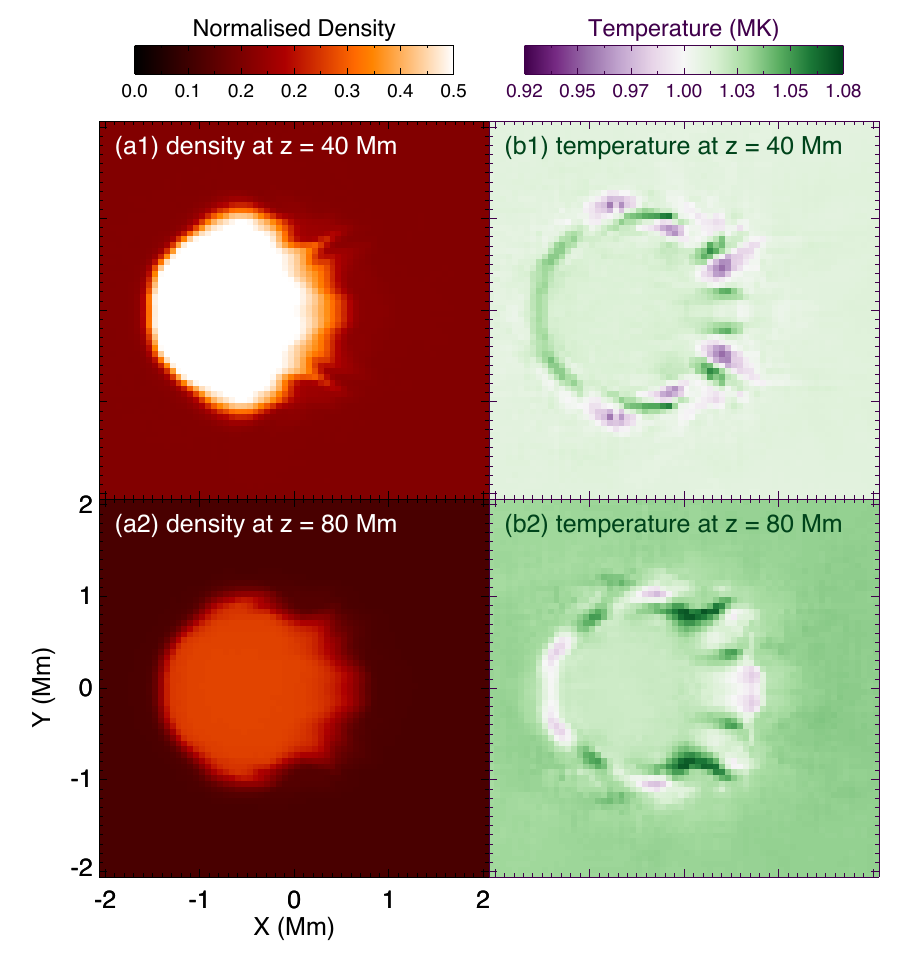}
    \caption{Cross-section profiles of (a) density and (b) temperature at $t=1140$ s. The upper and lower rows correspond to different heights ($z=40$ Mm and 80 Mm). A relevant animation is available in the online Journal.}
    \label{fig:5}
\end{figure}

\begin{figure*}
    \centering
    \includegraphics[width=1.0\textwidth]{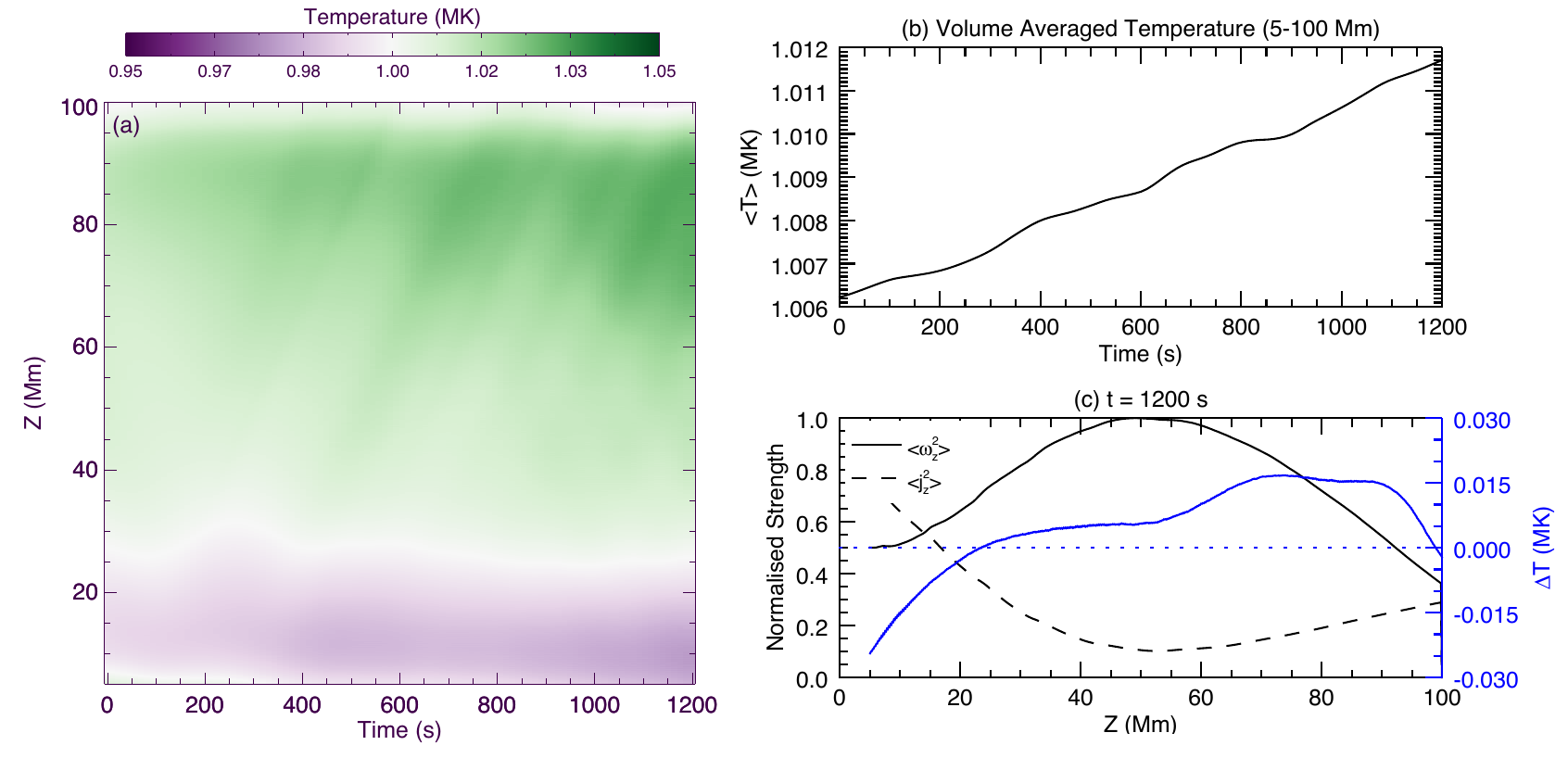}
    \caption{(a) Temporal and height variations of surface-averaged temperature. (b) Time series of the volume-averaged temperature within [-2, 2] Mm $\times$ [-2, 2] Mm $\times$ [5, 100] Mm. (c) The black solid and dashed lines depict the height dependence of normalised $\langle\omega_z^2\rangle$ and $\langle j_z^2\rangle$, respectively, which are averaged at each height over the time range from $t=0$ to 1200 s. The blue solid line represents the increase in surface-averaged temperature ($\Delta T$) relative to $t=0$ as a function of heights, with the blue dotted line corresponding to $\Delta T=0$.}
    \label{fig:6}
\end{figure*}

\begin{figure*}
    \centering
    \includegraphics[width=1.0\textwidth]{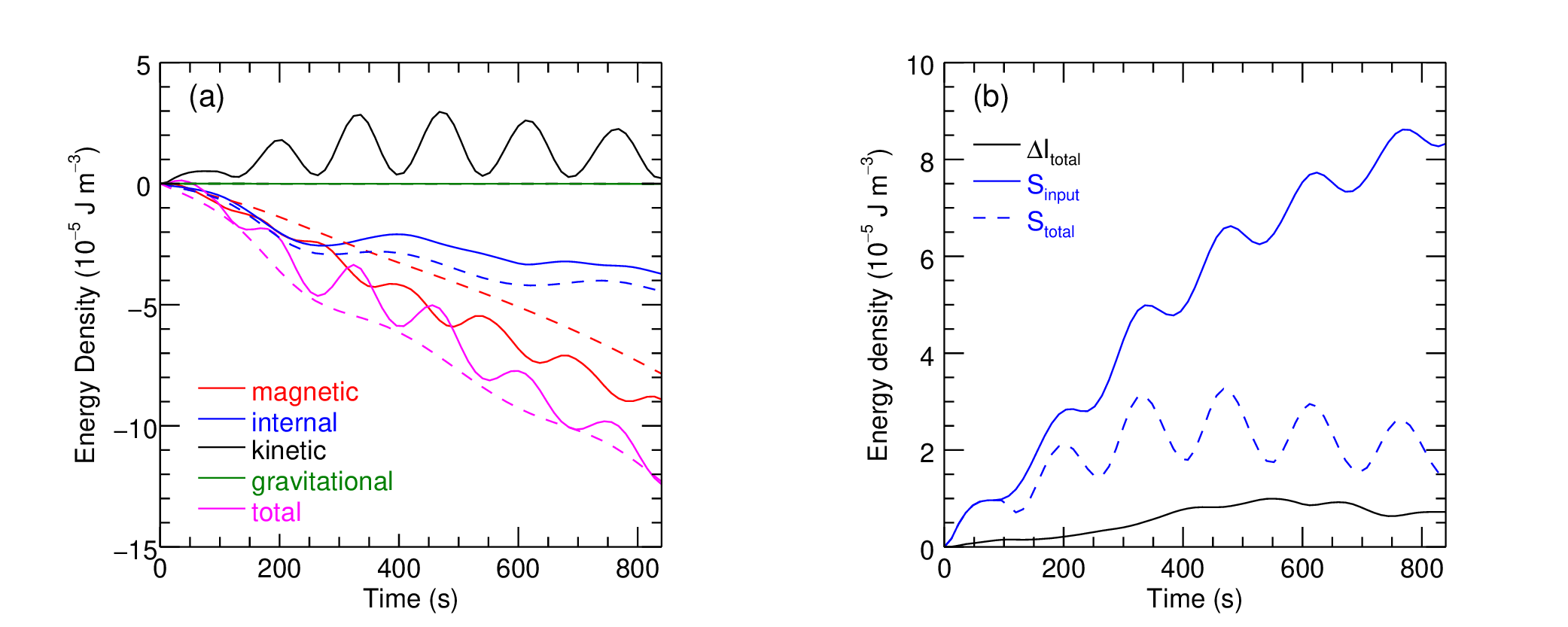}
    \caption{(a) Time series for the magnetic (red), internal (blue), kinetic (black), gravitational (green), and total (magenta) energy density relative to $t=0$. Solid and dashed lines represent simulation runs with and without a wave driver, respectively. (b) Time series of increased internal energy due to wave ($\Delta I_\mathrm{total}$, the black solid line), total input energy from the wave driver ($S_\mathrm{input}$, the blue solid line), and the ``trapped" portion of input energy ($S_\mathrm{total}$, the blue dashed line).}
    \label{fig:7}
\end{figure*}

In contrast to kink waves, the azimuthal Alfv\'{e}n waves generated by resonant absorption have a non-collective characteristic, which means that they can propagate at multiple magnetic surfaces each with a different Alfv\'{e}n speed \citep[e.g.,][]{terradas2010,soler2015}{}. Thus, phase mixing can form at the boundary layer, manifested as a deformed wavefront \citep{Heyvaerts1983}. The joint effect of resonant absorption and phase mixing leads to the small-scale structures appearing in the tube boundary in Figure \ref{fig:4}(b2) and (b3).

To further illustrate these small-scale structures, we plot the cross-section profiles of density and temperature at $z=$40 Mm and 80 Mm at $t=1140$ s, as shown in Figure \ref{fig:5}. We can see some finger-like density structures in Figure \ref{fig:5}(a1) and (a2), suggesting the initiation of KHI due to velocity shearing related to phase mixing. However, from the relevant animations, the development of KHI seems to be suppressed to a limited level. 
In contrast, for standing kink waves, the KHI can fully develop after resonant absorption and phase mixing, eventually leading to the formation of a turbulent state \citep[e.g.,][]{terradas2008,antolin2014,antolin2015,antolin2016,okamoto2015,howson2017,kara2017,kara2019,guo2019,guo2020,guo2023flux,shi2021a,shi2021b}. Specifically, the KHI vortices, which are a natural result for standing kink waves, cannot be generated in our simulation (for further discussion see Section \ref{subsec:KHI}).

As indicated in the previous studies \citep[][]{Pagano2017,pagano2018,pagano2020}, although KHI-induced turbulent vortices and non-linear state may not exist, phase mixing can still lead to some energy dissipation and heating by generating smaller structures. 
In Figure \ref{fig:5}(b), we can notice that the most significant temperature variations occur in the boundary layer, suggesting a direct association with resonant absorption and phase mixing here. However, there are both regions where temperature increases and decreases (compared to the initial uniform temperature of $\sim1$ MK), which could be interpreted as adiabatic heating (cooling) instead of real dissipation (see e.g., \citealt{kara2017,guo2019}). 

We then calculate the surface averaged temperature at every height and every time step. During the calculation, we restrict the region to be -2 to 2 Mm in both the $x$ and $y$ directions. Moreover, the oscillating flux tube does not go beyond this sub-region. The results are depicted in Figure \ref{fig:6}(a). Considering that there might be some numerical issues near the lower boundary ($z=0$ Mm), here we only focus on the part above 5 Mm. The temperature below 25 Mm decreases with time, while a temperature increase exists in parts above 25 Mm (also the main part of the flux tube). Furthermore, the amount of temperature change ($\Delta T$) also increases with height (see also the blue solid line in Figure \ref{fig:6}(c)). In Figure \ref{fig:6}(b), we plot time series of temperature averaged in the volume from $z=5$ Mm to 100 Mm, which also shows a growing trend, rising by approximately 6000 K within 20 min.

There may be two main factors that contribute to the increase in average temperature: viscous dissipation and Ohmic/resistivity dissipation. We calculate the values of $\langle\omega_z^2\rangle$ and $\langle j_z^2\rangle$ by averaging them at each height over the time range from $t=0$ to 1200 s, and plot them (after normalisation) as a function of height $z$ in Figure \ref{fig:6}(c). They can be compared with the temperature change $\Delta T$ during the same time range. 
Roughly speaking, the heating effect appears to be more pronounced at the heights where the vorticity is relatively larger and the current density is relatively lower. 
Therefore, for propagating kink waves, viscous dissipation might be more efficient in converting wave energy into thermal energy. 
Furthermore, our numerical setup introduces a quite large numerical resistivity (see Section \ref{sec:method}), but the spatial distributions of $\langle j_z^2\rangle$ and $\Delta T$ barely show any correlation, which further suggests that resistivity dissipation may not be dominant.
In comparison, heating induced by standing modes inside closed coronal loops exhibits a clear preference at lower heights (near footpoints) where resistivity dissipation is more significant, while viscosity dissipation dominates near the loop apex (see \citealt{tvd2007,antolin2014,kara2017,guo2019,shi2021a,shi2021b}). Further understanding of the contributions from these two dissipation mechanisms for propagating kink waves is important but beyond the scope of our current study.

\subsection{Analysis of energy density}\label{subsec:energy}

The analysis of volume averaged energy density can reveal valuable information of the system \citep[e.g.,][]{kara2019,guo2019,shi2021a,shi2021b}. We estimate the energy density inside the same volume ($V_0$) as that in Figure \ref{fig:6}(b), namely, [-2, 2] Mm $\times$ [-2, 2] Mm $\times$ [5, 100] Mm. Following \cite{kara2019}, the kinetic ($K(t)$), magnetic ($M(t)$), internal ($I(t)$), and gravitational ($G(t)$) energy density variations are given by:
\begin{equation}
    K(t)=\frac{1}{V_0}\int_{V_0}\frac{\rho(t)v(t)^2}{2}\mathrm{d}V-\frac{1}{V_0}\int_{V_0}\frac{\rho(0)v(0)^2}{2}\mathrm{d}V\,,
\end{equation}
\begin{equation}
    M(t)=\frac{1}{V_0}\int_{V_0}\frac{B(t)^2}{2\mu_0}\mathrm{d}V-\frac{1}{V_0}\int_{V_0}\frac{B(0)^2}{2\mu_0}\mathrm{d}V\,,
\end{equation}
\begin{equation}
    I(t)=\frac{1}{V_0}\int_{V_0}\frac{p(t)}{\gamma-1}\mathrm{d}V-\frac{1}{V_0}\int_{V_0}\frac{p(0)}{\gamma-1}\mathrm{d}V\,,
\end{equation}
\begin{equation}
    G(t)=\frac{1}{V_0}\int_{V_0}\rho(t)\Phi(t)\mathrm{d}V-\frac{1}{V_0}\int_{V_0}\rho(0)\Phi(0)\mathrm{d}V\,,
\end{equation}
where $\gamma=5/3$ is the adiabatic index, and $\Phi$ denotes the gravitational potential. The total energy density variation is 
\begin{equation}
    E(t)=K(t)+M(t)+I(t)+G(t)\,.
\end{equation}
For comparison, we also perform a new run without the footpoint driver. The time series of energy density for both runs are shown in Figure \ref{fig:7}(a). During the no-driver run (indicated by dashed lines), the total energy experiences a decrease with time, because the energy can be lost from the boundaries, especially from the upper boundary which is set as an extended outflow boundary with artificial velocity absorption (see Section \ref{sec:method} and Figure \ref{fig:2}). However, this can be acceptable because plumes in the coronal hole are naturally associated with mass and energy loss caused by the solar wind \citep[e.g.,][]{Poletto2015,morton2015,tian2021}.

For the run with a driver (marked by solid lines), we can notice an increase in kinetic energy relative to the no-driver case, which is obviously due to the input of Poynting flux carried by the driver. However, the magnetic energy decreases faster than that of the no-driver case. This could be a result of increased lateral energy leakage through side boundaries after kink waves are excited \citep[see e.g.,][]{guo2019,shi2021b}. The internal energy has a slight increase compared to the no-driver case, suggesting that the propagating kink wave can indeed dissipate the kinetic energy and lead to an increase in the internal energy of the system.

In Figure \ref{fig:7}(b), we plot the input energy density $S_\mathrm{input}(t)$ inside $V_0$ with a blue solid line, which is estimated by 
\begin{equation}
    S_\mathrm{input}(t)=\frac{1}{V_0}\int_0^t\int_{A_\mathrm{b}}(\vec{v}\times\vec{B})\times\vec{B}\cdot\mathrm{d}\vec{A'}\mathrm{d}t'\,,
\end{equation}
where $A_\mathrm{b}$ denotes the area of the bottom boundary. However, not all of the input energy (Poynting flux) can remain inside the volume $V_0$. As mentioned above, there is also energy flux escaping through the upper boundary. Thus, we also calculate how much energy can be ``trapped" inside $V_0$, given by
\begin{equation}
    S_\mathrm{total}(t)=\frac{1}{V_0}\int_0^t\int_{A}(\vec{v}\times\vec{B})\times\vec{B}\cdot\mathrm{d}\vec{A'}\mathrm{d}t'\,,
\end{equation}
where $A$ includes all six outer surfaces of $V_0$. The evolution of $S_\mathrm{total}(t)$ is depicted by the blue dashed line in Figure \ref{fig:7}(b), and it approximately saturates at around $2\times10^{-5}\,\text{J m}^{-3}$. This trapped energy can partially convert to internal energy, manifesting as the internal energy increase $\Delta I$ (relative to the no-driver case, shown by the black line in Figure \ref{fig:7}(b)).

\section{Forward modelling with FoMo}\label{sec:fomo}

\begin{figure*}
    \centering
    \includegraphics[width=1.0\textwidth]{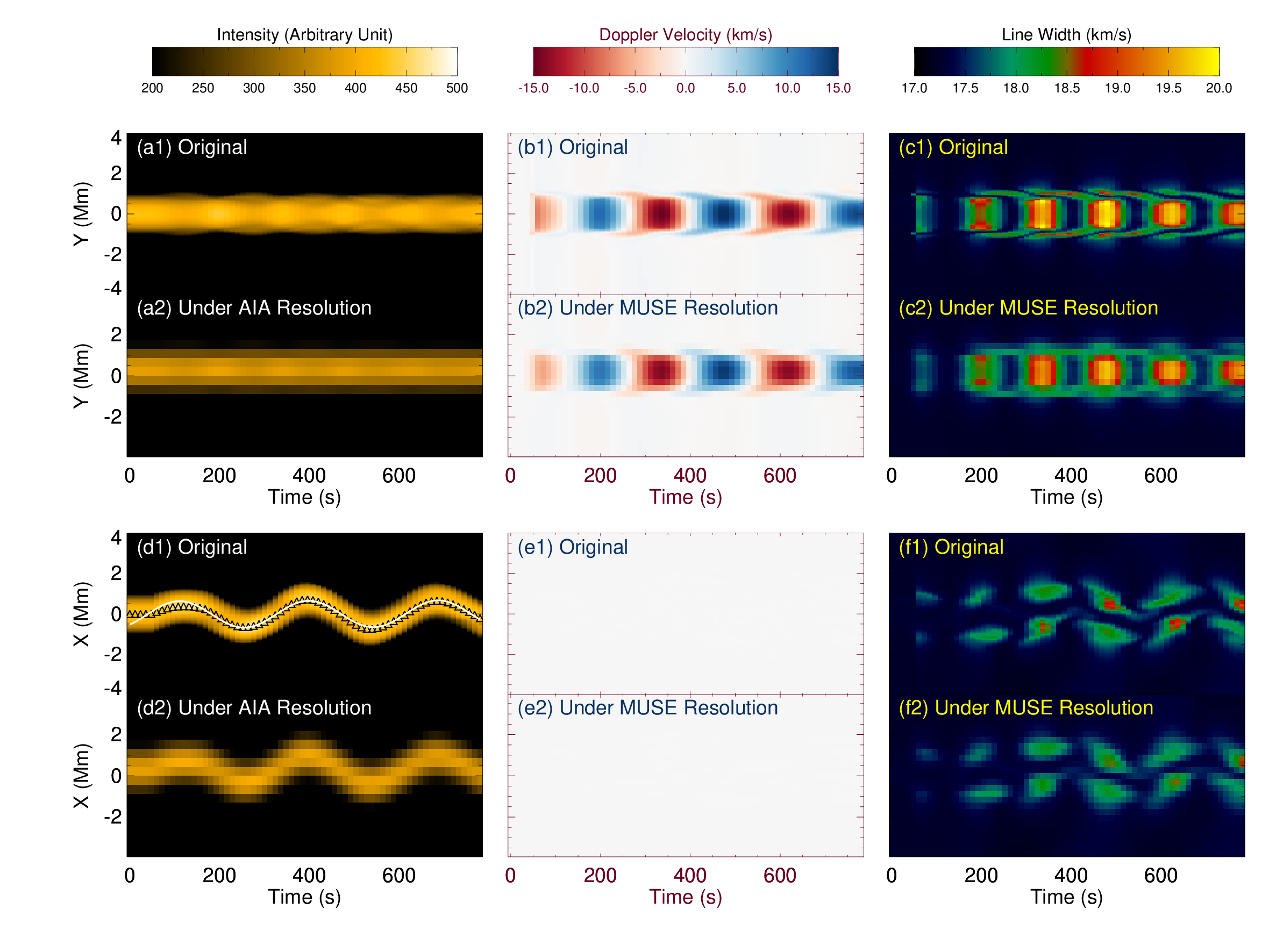}
    \caption{Time-distance maps of forward modelling results at $z=40$ Mm. The upper and lower rows correspond to the LOS view and POS view (defined in Section \ref{sec:fomo}), respectively. (a) TD maps of Fe \textsc{ix} 17.1 nm line intensity with the original spatial resolution and a degraded resolution comparable to AIA. (b-c) TD maps of Doppler velocity and line width with the original spatial resolution and a degraded resolution comparable to MUSE. Panels (d)-(f) are similar but for the POS view. Panel (d1) also demonstrates how we obtain the wave parameters from intensity TD maps. The tube centre positions are marked with black triangles, and the sinusoidal fitting results are depicted with a white curve.}
    \label{fig:8}
\end{figure*}

\begin{figure}
    \centering
    \includegraphics[width=0.45\textwidth]{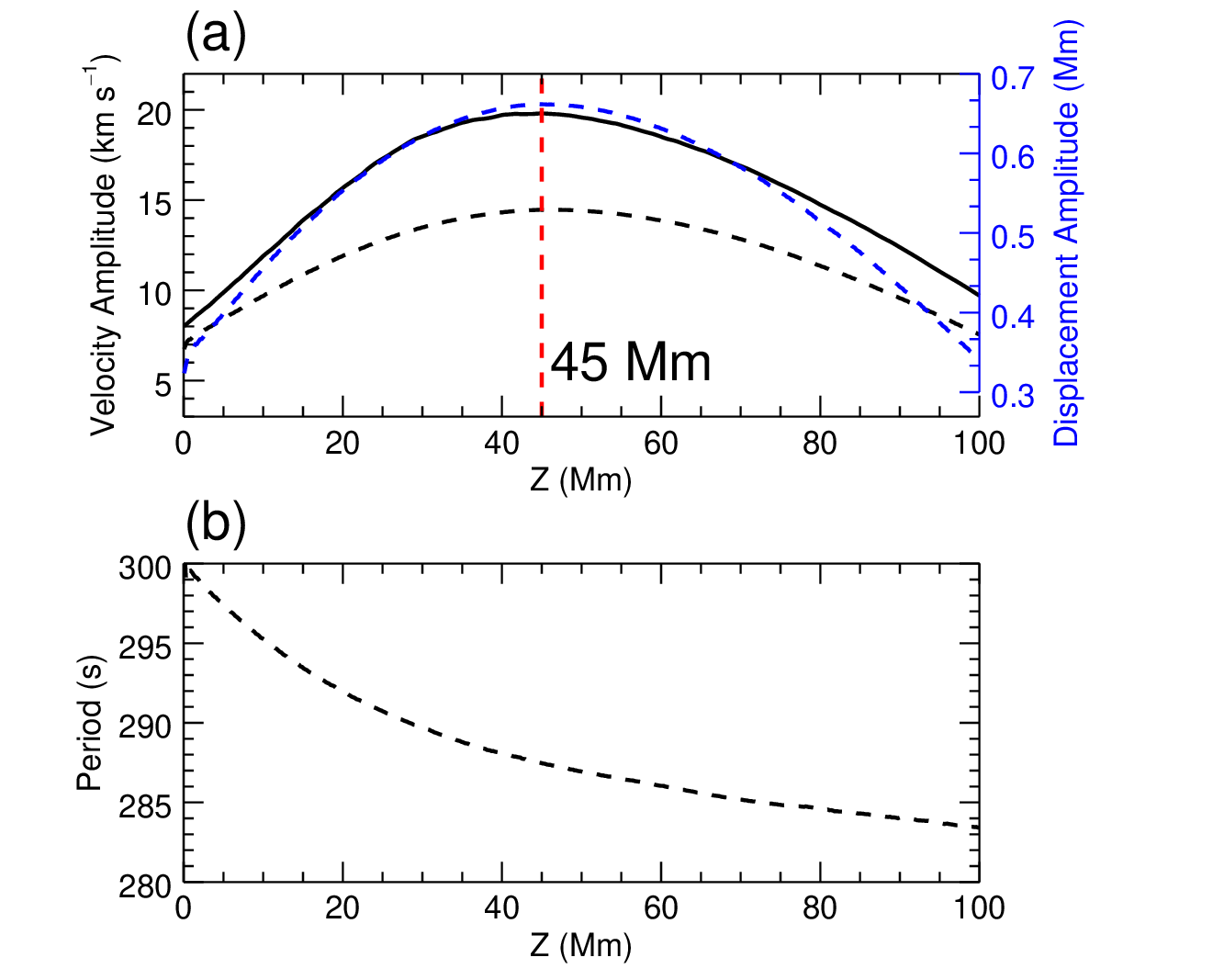}
    \caption{(a) Height variation of the velocity (black) and displacement (blue) amplitude, with the red dashed line indicating the position where the amplitude reaches the maximum ($z=45$ Mm). (b) Height variation of the wave period. 
    The black solid line is extracted from the original simulation output data, and all other dashed lines in both panels are obtained with forward modelling results as described in Section \ref{subsec:z_variation}.}
    \label{fig:9}
\end{figure}

It is of interest to investigate the observational properties of propagating kink waves in our simulation. Here we employ the FoMo code \citep{tvd2016} to forward-model the numerical results. The Fe \textsc{ix} 17.1 nm line is utilised because it is a typical coronal line in studying propagating transverse waves \cite[e.g.,][]{McIntosh2011,thurgood2014,morton2015,weberg2018,weberg2020}. For every snapshot (time step), we can obtain the synthesised spectra as a data cube with three dimensions: new $x$ or $y$ (depending on different LOS), new $z$, and wavelength $\lambda$. Then we perform a single Gaussian fit to the line profiles at each pixel to obtain the 2D map of line intensity, Doppler velocity and line width. In Figure \ref{fig:8}, we plot TD maps for these parameters at $z=40$ Mm. 

We choose two different LOS angles to do the forward modelling. The first one is parallel to the $x$ axis (shown in Figure \ref{fig:8}(a-c)), with the dominant velocity perturbation along the LOS; and the second one is parallel to the $y$ axis (shown in Figure \ref{fig:8}(d-e)), with the dominant velocity perturbation perpendicular to LOS (or in the POS). Therefore, we name these two cases as LOS view and POS view.

In order to directly compare FoMo results with observations, we also degrade the original spatial resolution with the method described in \cite{chen2021}. For the line intensity, the resolution is degraded to match SDO/AIA EUV imaging observations, which have a pixel size of $0.6''$ ($\sim$0.44 Mm) and a spatial resolution of $1.5''$ \citep{lemen2012}. Meanwhile, the Doppler velocity and line width can only be obtained by coronal EUV spectrometers. Although the Fe \textsc{ix} 17.1 nm line is not included in the coronal spectrometers currently at work,
the Multi-slit Solar Explorer \citep[MUSE;][]{DePontieu2020}, a proposed mission scheduled for launch in 2027, will have the capability to acquire high-resolution 17.1 nm spectra. Hence, here we degrade the original maps of Doppler velocity and line width to the spatial resolution ($0.4''$) of MUSE. Finally, we extract our simulation outputs every 12 s, matching the cadence achievable by both AIA and MUSE \citep{lemen2012,DePontieu2020,dePontieu2022}. 


\subsection{LOS view}\label{subsec:LOS}

We first focus on the results of the LOS view. As expected, we cannot see any transverse displacements from the original intensity TD maps (Figure \ref{fig:8}(a1)). Interestingly, there are some variations in the tube width and intensity with time, which have been usually interpreted as longitudinal waves such as sausage waves before
(e.g., \citealt{morton2012,tian2016}; see also the review by \citealt{libo2020}). However, these features are actually related to the cross-section deformation of the flux tube when it oscillates transversely (that is, moving forward and backward periodically for this view). For example, when the flux tube returns to its initial position, the velocity perturbation reaches the largest, causing the cross-section to stretch from a circle to an elliptical shape, with a slightly reduced span perpendicular to the velocity (see the online animation associated with Figure \ref{fig:5}). It further leads to the change of integrated optically thin emissions. Similar scenarios can be found in \cite{Cooper2003} and \cite{yuan2016} for standing kink waves. 
Another potential explanation is associated with the first fluting mode generated by nonlinearity \citep[see][]{ruderman2010}.
After degrading to AIA resolution (Figure \ref{fig:8}(a2)), the change of tube width cannot be captured, although some signatures of periodic intensity oscillation still remain, with a period that is approximately half of the wave period. 

For standing kink waves in closed coronal loops, forward modelling reveals that some apparent strands along the loop can be created as an LOS effect of KHI, offering a potential approach to observe this instability in coronal loops \citep{antolin2014}. Conversely, here we cannot produce such strand-like features, possibly due to the insufficient development of KHI for propagating modes as described in Section \ref{subsec:heating}. However, the periodic brightening observed in the intensity time-distance (TD) maps, as shown in Figure \ref{fig:8}(a2), may provide a novel way to indirectly observe signatures of propagating kink waves when the velocity perturbation is in the LOS. The result also suggests that detecting such signals using AIA imaging data could be achievable, and it may be even easier with higher resolution observations from the Extreme Ultraviolet Imager (EUI; \citealt{rochus2020}) on board
the Solar Orbiter (SolO; \citealt{muller2020}). Furthermore, we can easily distinguish such kink-wave-induced intensity variations from longitudinal waves \citep[e.g.,][]{Meadowcroft2024} because their propagating speeds differ significantly; the former would be one or two orders of magnitude larger than the latter \citep{morton2015,liu2015,Jess2016}.  As a result, it appears promising to detect such signatures with SDO/AIA or SolO/EUI in the near future.

From the TD map of Doppler velocity shown in Figure \ref{fig:8}(b), we can clearly see the kink wave signals presenting as the 
alternately generating red and blue shifts, as well as the phase mixing in the boundary layer. Such a pattern can be also found in the case of standing waves (\citealt{antolin2017,guo2019}). Additionally, the boundary characteristics can be detected under the MUSE resolution, which suggests that MUSE will have the capability to directly detect the signals of resonant absorption and phase mixing of propagating kink waves. Furthermore, the fine structures at the tube boundary are also captured by the TD map of line widths from Figure \ref{fig:8}(c). Considering that the temperature across the flux tube is close to 1 MK, the thermal width can be estimated by $\sigma_\mathrm{T}=\sqrt{2k_\mathrm{b}T/m_\mathrm{ion}}\sim17\,\text{km s}^{-1}$, where $k_\mathrm{b}$ and $m_\mathrm{ion}$ are the Boltzmann constant and mass of the ion Fe \textsc{ix}, respectively. Hence, the pattern presented here reflects the temporal and spatial variation of non-thermal broadening or non-thermal velocity ($v_\mathrm{nth}$), which can be naturally enhanced by generation of small scale structures at the tube boudary.

Another interesting feature is that the line width fluctuates in the tube centre with about half the wave period ($\sim$150 s). The largest line width ($\sim20$ km s$^{-1}$) gives a non-thermal velocity of $v_\mathrm{nth}\sim\sqrt{20^2-17^2}\text{ km s}^{-1}=10.5\text{ km s}^{-1}$.
This feature will be further discussed in the next subsection. 

\subsection{POS view}\label{subsec:POS}

The forward-modelling results of the POS view are relatively simple. The intensity, as shown in Figure \ref{fig:8}(d), oscillates transversely and sinusoidally over time, similar to previous observational studies with AIA \citep{McIntosh2011,thurgood2014,morton2015,weberg2018,weberg2020}. Also, as expected, no apparent Doppler velocity signals can be seen in Figure \ref{fig:8}(e). 
However, in the realistic case, when there is an inclination angle between the line of sight and the direction of this view, there will be observable Doppler signals (as a projection of main velocity perturbations). Such a case has been investigated in standing wave simulations such as \cite{antolin2015,antolin2017} and \cite{guo2019}. 

As for the line width shown in Figure \ref{fig:8}(f), the appearance of intensified non-thermal broadening is related to the formation of small-scale structures by phase mixing in the boundary layer. However, the development of these signals quickly saturates, suggesting that the phase-mixing-induced KHI will not fully develop into a turbulent state, consistent with the results presented in Section \ref{subsec:heating} and relevant animations of Figure \ref{fig:5}. Moreover, the non-thermal broadening exhibits a periodic variation that is simultaneous with that in the tube centre of the LOS view, and thus provide an interpretation of these patterns in Figure \ref{fig:8}(c). Specifically, the latter can be regarded as the result of a superposition of the former at both boundaries.
Finally, the degraded TD map (panel f2) illustrates that MUSE might have sufficient spatial and temporal resolutions to capture these localised fine structures. However, the total line width ($\sim$18.5 km s$^{-1}$) is only minorly
larger than the thermal width ($\sim$17 km s$^{-1}$), which appears likely to be covered by the instrumental broadening and noise, and thus may not be resolved in real observations.

Although the non-thermal velocities we obtained here appears to be too small to be detected for a single flux tube, we can expect that when there are multiple tubes in the line of sight, the integration effect will generate a much larger $v_\mathrm{nth}$ that is comparable to the results of spectral observation \citep[$\sim$20-50 km s$^{-1}$; see e.g.,][]{tvd2008,banerjee2009,Hahn2012,bemporad2012,hahn2013,morton2015,zhu2021,zhu2024}. Most previous studies tend to interpret observed non-thermal line broadening as a result of torsional Alfv\'{e}n waves or Alfv\'{e}n wave turbulence. However, our findings suggest that the LOS superposition of multiple flux tubes carrying kink waves can also generate the observed $v_\mathrm{nth}$, further supporting the conclusion in previous studies \citep[][]{McIntosh2012,pant2019,pant2020,Fyfe2021}. 

\subsection{Altitude variations of wave parameters}\label{subsec:z_variation}

In imaging observations, there is a set of techniques widely used to estimate wave parameters including amplitude and period from intensity TD maps. Specifically, the first step is to determine the tube centre with a Gaussian fitting \citep[e.g.,][]{thurgood2014,anfin2015} or some empirical thresholds of intensity spatial gradient \citep[e.g.,][]{weberg2018,weberg2020}. With the time series of tube centre positions (i.e., displacements), we can perform fitting with a sine function \citep[e.g.,][]{thurgood2014} or conduct Fourier transformation \citep[e.g.,][]{Duckenfield2018,weberg2018,weberg2020,bate2022,zhong2023} to obtain the displacement amplitude $\xi$ and period $P$. Subsequently, the velocity amplitude can be calculated with 
\begin{equation}\label{eq:v_a}
    v_\mathrm{amp}=2\pi \xi/P\,.
\end{equation} 
Here we adopt the similar analysis techniques described in \cite{gao2022} and \cite{gao2024} with intensity TD maps at every height from 0 to 100 Mm, to obtain the wave amplitude and period as a function of height ($z$). An example is shown in Figure \ref{fig:8}(d1) with the tube centre positions marked by black triangles and the fitted sinusoidal curve marked by the white solid line. 

The altitude variation of these wave parameters is presented in Figure \ref{fig:9}. The dashed lines in panel (a) are amplitudes acquired from the TD maps based on FoMo results, with the black (blue) one indicating the velocity (displacement) amplitude; while the solid line, as described in Section \ref{subsec:amplitude}, is the velocity amplitude calculated with original simulation output data. All of them exhibit a similar trend with height, peaking at about 45 Mm. However, the difference between the black solid and dashed lines implies that the velocity amplitude obtained from imaging observations using Equation \ref{eq:v_a} may potentially underestimate the amplitude and, consequently, the associated energy flux. Another plausible explanation for this discrepancy could be some averaging and line-of-sight integration effects, as we only use $v_x$ at the $z$ axis to obtain the amplitude corresponding to the black solid line.

From Figure \ref{fig:9}(b), it appears that the period undergoes a slight but consistent decrease with height, ranging from 300 s at $z=0$ Mm to 283 s at $z=100$ Mm. This trend is likely attributed to errors stemming from the fitting method when there is a non-oscillatory stage in TD maps (as shown in Figure \ref{fig:8}(d)). This stage exists because the wavefront requires time to propagate to a specific height. As the height increases, such a stage persists for a longer duration, resulting in an underestimation of the period. 
When extending the time series for fitting (e.g., to 1200 s), the period at all heights can return to a value very close to the driver period (300 s), which is expected. 
However, here we choose to present forward modelling results and conduct fitting for only a relatively short duration (approximately 2.5 oscillation cycles), because in observations it is rare to detect propagating kink wave events with more cycles \citep{weberg2018,weberg2020}. Thus, Figure \ref{fig:9}(b) provides valuable insights for future observational studies, indicating the importance of carefully removing the non-oscillatory part before extracting wave parameters, particularly the periods. Given that wave periods play an important role in coronal seismology (e.g., \citealt{naka2001,zhong2023NatSR}) and energy estimations (e.g., \citealt{bate2022,petrova2023,lim2023}), a thorough investigation on how analysis techniques affect period measurement is desired in the future.

\section{Discussion}\label{sec:discussion}

\subsection{Implications for coronal seismology of relative density profiles}\label{subsec:seis}

\begin{figure*}
    \centering
    \includegraphics[width=1.0\textwidth]{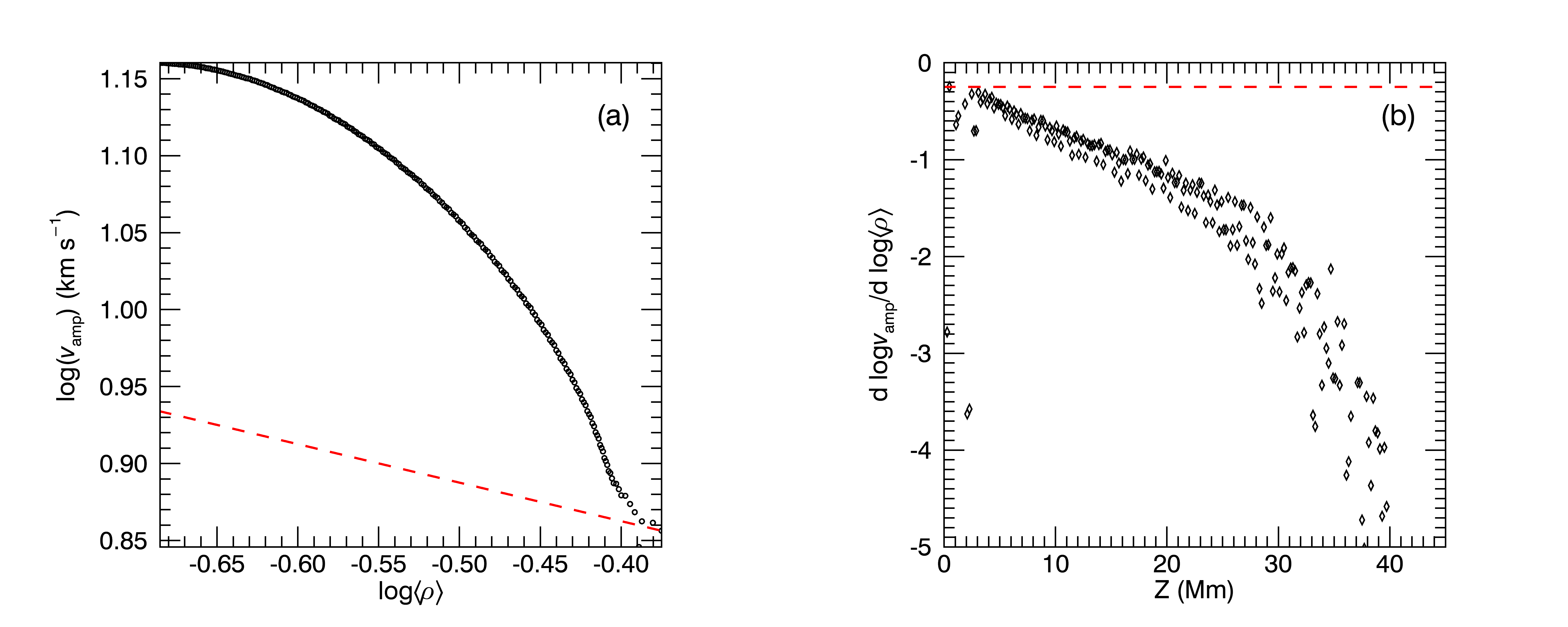}
    \caption{(a) Scatter plot between $\log\langle\rho\rangle$ and $\log v_\mathrm{amp}$. The red dashed line indicates their theoretical relation with a slope of $-1/4$ based on Equation (\ref{eq:seis}). (b) The derivative $\mathrm{d}\log v_\mathrm{amp}/\mathrm{d}\log\langle\rho\rangle$ as a function of height ($z$). The theoretical value ($-1/4$) based on Equation (\ref{eq:seis}) is also depicted with the red dashed line. We note that in panel (a), the height decreases from left to right.}
    \label{fig:10}
\end{figure*}

For propagating kink waves, the velocity amplitude can be expressed by \citep[e.g.,][]{soler2011,morton2012,weberg2020}
\begin{equation}\label{eq:va_theory}
    v_\mathrm{amp}(z)=C\sqrt{c_\mathrm{k}(z)}R(z)\,,
\end{equation}
where $C$ is a constant, and $R(z)$ is the radius of the flux tube at the height of $z$, which is also nearly constant for our model. Another simplified parameter is the magnetic field $B$, which varies very little in the whole simulation domain (see Figure \ref{fig:1}(c)). Thus, the kink speed $c_\mathrm{k}(z)$ is proportional to $1/\sqrt{\langle\rho\rangle}$, and finally we have 
\begin{equation}\label{eq:seis}
   v_\mathrm{amp}\propto \langle\rho\rangle^{-1/4}\,.
\end{equation}
Such a relation is utilised by some observational studies to derive the relative density profile as a function of height in the corona, especially the open field region (\citealt{morton2015,weberg2020}). Similar seismology methods have also been used in other structures such as spicules \citep{verth2011}, jets \citep{morton2012}, chromospheric mottles \citep{Kuridze2013}, and fibrils \citep{morton2014}. However, Equation (\ref{eq:seis}) does not consider the variation of the amplitude due to the combined effects of resonant absorption and the stratification (see Section \ref{subsec:amplitude}). Therefore, here we assess the seismology method with the simulation data to further investigate the effect of resonant absorption.

In Figure \ref{fig:10}(a), we plot the relation between velocity amplitude (derived based on forward-modelling result with Equation \ref{eq:v_a}) and average density at the same height in the log-log scale. We only select the data samples below 45 Mm, where $v_\mathrm{amp}$ increases with height. It is shown that there is no linear correlation with a slope of -1/4 between these two values (as marked by the red dashed line). Equation (\ref{eq:seis}) can be further examined by calculating $\dfrac{\mathrm{d}\log v_\mathrm{amp}}{\mathrm{d}\log\langle\rho\rangle}$ (the slope in Figure \ref{fig:10}(a)) as a function of height, which is shown in Figure \ref{fig:10}(b). The theoretical value without considering resonant damping should be -1/4, and it is indicated by the red dashed line. We can notice that, aside from very low altitudes where resonant absorption may not be well developed (see also Figure \ref{fig:4}(b1)), the calculated derivative values generally fall below the theoretical value of -1/4, with the deviation increasing at higher altitudes. Hence, if we still use Equation (\ref{eq:seis}) to derive the relative density change, there will be an underestimation of density.

It is expected that such a result is associated with the resonant damping rate, which can be quantified by the equilibrium parameter $\xi_\mathrm{E}$:
\begin{equation}\label{eq:quality_factor}
    \xi_\mathrm{E}=\frac{2}{\pi}\frac{R}{l}\frac{\rho_i+\rho_e}{\rho_i-\rho_e}=\frac{2}{\pi}\frac{R}{l}\left(1+\frac{2}{\zeta-1}\right)\,.
\end{equation}
Then the damping length is:
\begin{equation}
    L_\mathrm{D}=c_\mathrm{k}\xi_\mathrm{E}P=\xi_\mathrm{E}\lambda\,,
\end{equation}
where $\lambda$ is the wavelength. To validate our findings, we perform two additional runs: one with a larger $\xi_\mathrm{E}$ (achieved by reducing the density contrast $\zeta$ from 3 to 2) and another with a longer wavelength (achieved by increasing the period $P$ from 300 s to 900 s) to investigate cases with larger 
$L_\mathrm{D}$. The results demonstrate that the deviation from -1/4 remains non-negligible for both runs. As a consequence, our simulation illustrates that, when conducting such seismological diagnostics, resonant absorption should be taken into account even if we observe an increasing amplitude with height as found in e.g., \cite{morton2015} and \cite{weberg2020}. Specifically, the resonant damping could lead to an underestimation of the average density when applying Equation (\ref{eq:seis}). Moreover, according to Equation (\ref{eq:seis}), the density is proportional to the fourth power of the velocity amplitude, which can significantly amplify any error brought from $v_\mathrm{amp}$, further stressing the lack of robustness of this seismology method.

The conclusion drawn here may provide an additional explanation for the results in \cite{weberg2020}. They applied Equation \ref{eq:quality_factor} and observed $v_\text{amp}$ as a function of height to derive the relative density profile above the solar limb. In their Figure 7, a rapid decrease of the derived relative density between 5-10 Mm suggests the presence of an extended transition region up to this height. However, this is much larger than the classical transition region height of approximately 1-2 Mm (e.g., \citealt{Avrett2008}) above the solar surface. \cite{weberg2020} interpreted the discrepancy as a potential result of dynamic fibrils and spicules reaching high in the corona \citep[see e.g.,][]{dePontieu2007,pereira2012}. However, if resonant absorption is considered, the discrepancy can be attributed to the underestimation of density when using Equation (\ref{eq:seis}) in a quite straightforward way. Based on Figure \ref{fig:10}(a), when the velocity amplitude increases from $10^{0.88}$ ($\sim$7.6) km s$^{-1}$ to $10^{0.93}$ ($\sim$8.5) km s$^{-1}$, representing a 15\% increase, the derived density can show a decreasing by approximately 60\% if resonant damping is not considered. In contrast, if resonant damping is taken into account, the density only decreases by about 10\%. Therefore, neglecting damping can exaggerate the density gradient, leading to an overestimation of the transition region height.

Further consideration should be given to the limitations of our current model, particularly regarding the disparities between our model and the actual solar atmosphere. Our current simulation does not include the cross-section area expansion and magnetic field weakening with height (see e.g., \citealt{soler2011,ruderman2013,ruderman2019,lopin2017,Guo2024}) for simplification. However, the magnetic flux conservation gives that $B(z)R(z)^2$ is approximately a constant for a flux tube. Thus, according to Equation (\ref{eq:va_theory}) and $c_\mathrm{k}=B/\sqrt{\mu_0\langle\rho\rangle}$, the relation (\ref{eq:seis}) remains valid regardless of any height variations of $B(z)$ and $R(z)$ as long as the magnetic flux is conserved. Nevertheless, the expansion of the flux tube and longitudinal inhomogeneity of the magnetic field may influence the wave behaviour and strength of resonant absorption \citep[e.g.,][]{fedun2011,lopin2017,howson2019,soler2019,ruderman2019}, hence it would be interesting to investigate wave properties in an expanding magnetic flux tube with 3D MHD simulations in the future (currently, such works are mostly analytical). Recently, \cite{howson2019} developed a model for an expanding closed coronal flux tube, focusing on standing kink waves. However, in their model, the density remains uniform throughout the simulation domain. Thus, a more realistic MHD model, combining both density and magnetic field inhomogeneity, should be established and explored in future studies.

Another important seismological application of propagating kink waves is to diagnose the coronal Alfv\'{e}n speed and magnetic field by measuring its phase speed \citep{morton2015,long2017,magyar2018,yang2020,yang2020Sci,Baweja2024}. Our model is suitable for assessing such seismology techniques, which will be the topic of a forthcoming paper.

\subsection{Propagating kink waves and energy balance in coronal open field region}

As described in Section \ref{subsec:heating} and \ref{subsec:energy}, propagating kink waves might have a potential contribution to the energy balance in coronal open field regions, or coronal holes. According to the estimation in \cite{Withbroe1977}, the solar wind flux dominates the energy loss in the coronal hole, much larger than that of thermal conduction and radiative cooling. In our simulation, this energy loss can be roughly reflected in the decrease in energy density shown in Figure \ref{fig:7}(a), particularly for the no-driver case (dashed lines). For the case with a kink wave driver (see Figure \ref{fig:7}(b)), most input wave energy ($S_\mathrm{input}$) flows out from the upper boundary, while some energy ($S_\mathrm{total}$) can remain in the corona, as indicated by blue solid and dashed lines. The latter can be dissipated and partially transformed into internal energy, which may contribute to compensating the radiative loss and driving the solar wind.

There are two potential approaches to enhance wave heating efficiency to balance the energy loss (apart from the solar wind flux) in the coronal hole.
The first one is to input more energy at the lower boundary, which can be achieved by increasing the amplitude/frequency of the driver \citep[e.g.,][]{kara2019} and/or employing a broad-band driver instead of a mono-periodic one \citep[e.g.,][]{pascoe2015,magyar2017,magyar2018,pagano2019,pagano2020}. However, further investigation requires a detailed parameter survey, which exceeds the scope of this study. The second approach is to augment the dissipation rate of wave energy into internal energy, which is associated with the development of KHI and turbulence, and will be discussed in the subsequent subsection.

\subsection{KHI development of propagating waves}\label{subsec:KHI}

In our simulation, the efficiency of energy dissipation is constrained by the incomplete development of the KHI and relevant turbulence (see Section \ref{subsec:heating}). This is one of the primary distinctions between propagating and standing modes. It has been known since the early 1980s that propagating Alfv\'{e}n waves undergoing phase mixing are stable to KHI while standing waves are unstable \citep{Heyvaerts1983}. As indicated by \cite{zaqarashvili2015}, for propagating (coupled) Alfv\'{e}n waves induced via resonant absorption, the magnetic and velocity perturbations are out of phase. An external horizontal magnetic field parallel to the velocity shearing can increase the instability threshold and consequently stabilize the KHI \citep[e.g.,][]{bennett1999,LiXH2018}. Consequently, when the velocity perturbation attains its peak, the magnetic field perturbation also reaches its maximum but in the opposite direction, thereby suppressing KHI. Conversely, for standing modes, the velocity amplitude peaks at the antinodes (e.g., the loop apex for fundamental oscillations of a closed coronal loop), while the magnetic perturbation there is zero \citep[e.g.,][]{Zaqarashvili2003,guo2020}. Hence KHI can be easily induced, with its heating effect shown to sufficiently counterbalance radiative losses in closed coronal loops \citep{shi2021b}. 

In fact, most previous numerical studies on propagating kink waves, including ours, have not reported the (fully) development of KHI \citep{pascoe2010,pascoe2011,pascoe2012,zaqarashvili2015,Pagano2017,howson2020}. However, \cite{pagano2019} reported that KHI can be generated with propagating modes, with the flux tube significantly deformed. One possible explanation for this inconsistency is that they used the observed power spectrum \citep{tomczyk2009} to set the wave driver, resulting in a driver that is in a nearly turbulent state. In such cases, the cylindrical symmetry of the flux tube might be broken, and instabilities could be more likely to occur \citep[see][]{Meringolo2023}. Nonetheless, the findings of \cite{pagano2019} have put it under debate whether KHI can effectively form and develop for propagating kink waves \citep[see also Section 4.2 in][]{morton2023}.

In our simulation, the small-scale structures at the tube boundary (see Figure \ref{fig:5}) may also be linked to uniturbulence \citep[e.g.,][]{magyar2017,magyar2019,tvd2020,Ismayilli2022}, which describes the nonlinear self-deformation of propagating kink waves, generating a cascade to smaller scales. Accepting this interpretation, it would be interesting to compare the energy cascade rate with theoretical predictions (e.g., \citealt{tvd2020} but with modification to incorporate longitudinal stratification) in the future.

\section{Conclusions}\label{sec:conclusion}

In this study, we conduct a 3D MHD simulation to investigate the behavior of propagating kink waves within a gravitationally stratified open magnetic flux tube. With both transverse and longitudinal density inhomogeneity incorporated, we can investigate their joint effect on these widely spread wave phenomena. Particularly, the consideration of gravitational stratification is missing in most previous numerical studies of the same type \citep{pascoe2010,pascoe2011,pascoe2012,magyar2017,magyar2018,Pagano2017,pagano2019,pagano2020,howson2020}. Furthermore, we perform forward modelling with FoMo to synthesise observational properties. Such analyses have been extensively conducted for simulations of standing kink waves \citep[e.g.,][]{antolin2015,antolin2016,antolin2017,guo2019,shi2021a}, but have rarely been applied to propagating modes within an individual open flux tube.

Our main findings can be summarised as follows:

1. Resonant absorption and gravitational stratification both influence the altitude variation of wave amplitude. While stratification dominates at lower heights, resonant absorption becomes more prominent at higher altitudes. 
When diagnosing the relative density profile with Equation \ref{eq:seis}, resonant damping needs to be considered to avoid possible underestimation, even in the regime where resonant damping is not dominant.

2. In contrast to standing kink modes, for propagating waves the KHI development appears to be suppressed after resonant absorption and phase mixing. Nonetheless, fine structures can still be formed in the tube boundary, leading to energy cascading to smaller scales and a slight temperature enhancement, especially at higher altitudes. A major portion of the input wave energy exits from the upper boundary, corresponding to the flux carried by the solar wind, while the remaining energy causes an internal energy increase, potentially contributing to balancing the radiative losses in coronal holes.

3. Forward-modelling results highlight the promising potential of future high-resolution coronal imaging spectrometers (such as MUSE) in directly detecting signatures of resonant absorption and phase mixing. Moreover, transverse motions of the flux tube can produce periodical variations in TD maps of intensity and line width with half the wave period. Notably, the former could potentially be captured using current EUV imagers, such as SDO/AIA and SolO/EUI.

Future works will focus on enhancing the complexity of the model and the driver. This may involve extending the model to incorporate multiple flux tubes 
and/or the lower atmosphere. 
Additionally, the driver could be replaced with a broadband one to better mimic the actual transverse motions at the base of the corona. Within the coronal open field region, other topics of interest include the connection with the solar wind (mass transport and acceleration), propagating intensity disturbances, and the influence of background flow on transverse waves. 
These topics are also worth investigating using our current model (or its modified versions).

\begin{acknowledgement}

This work is supported by the National Natural Science Foundation of China grant 12425301 and the National Key R\&D Program of China No. 2021YFA1600500.
Y.G. acknowledges support from the China Scholarship Council under file No. 202206010018.
M.G. acknowledges support from the National Natural Science Foundation of China (NSFC, 12203030).
T.V.D was supported by the European Research Council (ERC) under the European Union's Horizon 2020 research and innovation programme (grant agreement No 724326), the C1 grant TRACEspace of Internal Funds KU Leuven, and a Senior Research Project (G088021N) of the FWO Vlaanderen. Furthermore, TVD received financial support from the Flemish Government under the long-term structural Methusalem funding program, project SOUL: Stellar evolution in full glory, grant METH/24/012 at KU Leuven. The research that led to these results was subsidised by the Belgian Federal Science Policy Office through the contract B2/223/P1/CLOSE-UP. It is also part of the DynaSun project and has thus received funding under the Horizon Europe programme of the European Union under grant agreement (no. 101131534). Views and opinions expressed are however those of the author(s) only and do not necessarily reflect those of the European Union and therefore the European Union cannot be held responsible for them. K.K. acknowledges support by an FWO (Fonds voor Wetenschappelijk Onderzoek – Vlaanderen) postdoctoral fellowship (1273221N).
\end{acknowledgement}

\bibliographystyle{aa} 
\bibliography{ref.bib} 

\end{document}